\documentclass[a4paper,fleqn,usenatbib]{mnras}

\usepackage[T1]{fontenc}
\usepackage{ae,aecompl}

\usepackage{graphicx}	
\usepackage{amsmath}	
\usepackage{amssymb}	

\title[Fluorine]{Fluorine in the solar neighborhood: modelling the Galactic thick and thin discs}

\author[Grisoni et al.]{V. Grisoni$^{1,2}$\thanks{E-mail: valeria.grisoni@sissa.it}, D. Romano$^3$, E. Spitoni$^4$, F. Matteucci$^{5, 6, 2}$, N. Ryde$^{7}$, H. J{\"o}nsson$^{8,7}$\\
 $^1$ SISSA, Via Bonomea 265, 34136 Trieste, Italy\\  
 $^2$ INAF, Osservatorio Astronomico di Trieste, via G.B. Tiepolo 11, I-34131, Trieste, Italy\\
 $^3$ INAF, Osservatorio di Astrofisica e Scienza dello Spazio, Via Gobetti 93/3, 40129 Bologna, Italy\\
 $^4$ Stellar Astrophysics Centre, Department of Physics and Astronomy, Aarhus University, Ny Munkegade 120, DK-8000 Aarhus C,\\ Denmark\\
 $^5$ Dipartimento di Fisica, Sezione di Astronomia, Universit\`a di Trieste, via G.B. Tiepolo 11, I-34131, Trieste, Italy \\
 $^6$ INFN, Sezione di Trieste, via Valerio 2, 34134 Trieste, Italy\\
 $^7$ Lund Observatory, Department of Astronomy and Theoretical Physics, Lund University, Box 43, SE-22100 Lund, Sweden\\
 $^8$ Materials Science and Applied Mathematics, Malm\"o University, SE-205 06 Malm\"o, Sweden
}

\begin{document}
\date{Accepted . ; in original form xxxx}

\pagerange{\pageref{firstpage}--\pageref{lastpage}} \pubyear{xxxx}

\maketitle

\label{firstpage}

\begin{abstract}
We investigate the evolution of the abundance of fluorine in the Milky Way thick and thin discs by means of detailed chemical evolution models compared with recent observational data. The chemical evolution models adopted here have already been shown to fit the observed abundance patterns of CNO and $\alpha$-elements as well as the metallicity distribution functions for the Galactic thick and thin disc stars. We apply them here to the study of the origin and evolution of fluorine, which is still a matter of debate. First, we study the importance of the various sites proposed for the production of fluorine. Then, we apply the reference models to follow the evolution of the two different Galactic components. We conclude that rotating massive stars are important producers of F and they can set a plateau in F abundance below [Fe/H]=-0.5 dex, though its existence for [Fe/H]<-1 has yet to be confimed by extensive observations of halo stars. In order to reproduce the F abundance increase in the discs at late times, instead, a contribution from lower mass stars -- single asymptotic giant branch stars and/or novae -- is required. The dichotomy between the thick and thin discs is more evident in the [F/O] vs. [O/H] plot than in the [F/Fe] vs. [Fe/H] one, and we confirm that the thick disc has evolved much faster than the thin disc, in agreement with findings from the abundance patterns of other chemical elements. 
\end{abstract}

\section{Introduction}

The study of the origin and evolution of fluorine still represents a hot topic in the field of Galactic Archaeology. The only stable isotope of fluorine is $^{19}$F and its production is related to the physical conditions in stars. In literature, several stellar sites have been proposed for the production of fluorine, which can be summarized as follows.
\begin{enumerate}
\item \textit{Asymptotic giant branch (AGB) stars}. In AGB stars, during the He-burning thermal pulses, the $^{14}$N that is synthesized in the hydrogen-burning CNO cycle can produce fluorine by means of a chain of reactions, involving also neutrons and protons (see Forestini et al. 1992; Jorissen et al. 1992; Abia et al. 2011; Gallino et al. 2010; Cristallo et al. 2014). Fluorine can be brought to the surface by the 3rd dredge-up and then it is expelled into the interstellar medium (ISM), by stellar winds or during the planetary nebula phase. Fluorine produced in this way would be a secondary element, with yields depending on the metallicity; the production of fluorine in metal-poor AGB stars has also a primary origin, depending on how $^{13}$C is produced (Jorissen et al. 1992; Forestini et al. 1992; Cristallo et al. 2014). At high temperatures in stellar interiors, fluorine can be destroyed by helium-nuclei or proton-capture reactions and converted into Ne; thus, the AGB stars that produce fluorine should be less massive than $\sim$4 M$_{\odot}$ preventing the temperatures of hot bottom burning (Lugaro et al. 2004; Karakas 2010; Cristallo et al. 2014). Observationally, it has been shown that AGB stars do contribute to fluorine (Jorissen et al. 1992; Abia et al. 2015, 2019).
\item \textit{Rapidly rotating massive stars}. Rapidly rotating massive stars can produce primary fluorine from $^{14}$N, through proton and $\alpha$ captures in the presence of $^{13}$C; $^{14}$N derives from reactions with $^{12}$C, which comes from He burning in the massive star itself, and it is then of primary origin (Guer{\c{c}}o et al. 2019a). This chain of reactions clearly happens also in the non-rotating case, but the available amount of CNO nuclei is too small to contribute significantly to fluorine production (Limongi \& Chieffi 2018).
\item \textit{Wolf-Rayet (W-R) stars}. Massive stars evolving as W-R stars have also been proposed as fluorine producers (Meynet \& Arnould 1993, 2000). Fluorine is produced in the convective core of W-R stars, during the core He-burning phase; these stars can experience very strong stellar winds, which can prevent the destruction of fluorine. Fluorine is produced from $^{14}$N, which is normally produced during the CNO cycle as a secondary element; however, in massive stars N can be produced as a primary element if they suffer strong rotation. The difference between primary and secondary N is important, because in the case of secondary N, also the F behaviour would follow that of a secondary element, depending on the original stellar metallicity. However, Palacios et al. (2005) questioned the contribution of W-R stars to fluorine and they showed that it can be significantly reduced, when rotation is included in the stellar models.
\item \textit{The $\nu$-process in core-collapse supernovae}. The $\nu$-process active in core-collapse supernovae has also been proposed for the production of fluorine (Woosley \& Haxton 1988, Kobayashi et al. 2011b). Even if the cross sections of neutrino-nucleus reactions are small, the large flux of neutrinos released during the core-collapse can turn $^{20}$Ne in the outer envelopes of the collapsing star into fluorine. Fluorine produced by this process would be of primary origin. In this case, there are still several uncertainties, depending on the considered neutrino flux and energy.
\item \textit{Novae}. Also novae can in principle produce fluorine (Jos{\'e} \& Hernanz 1998). In the case of classical novae, the mechanism involved in the fluorine production is the reaction chain $^{17}$O(p,$\gamma$)$^{18}$F(p,$\gamma$)$^{19}$Ne, with the short-lived, $\beta^+$-unstable nucleus $^{19}$Ne that can be transported by convection to the outer and cooler layers of the envelope, where it decays into $^{19}$F. However, the yields of fluorine from novae are still very uncertain. 
\end{enumerate}
From the point of view of Galactic chemical evolution models, the evolution of fluorine in the Milky Way and the role of the different fluorine producers have been investigated in previous works (e.g. Timmes et al. 1995; Meynet \& Arnould 2000; Renda et al. 2004; Kobayashi et al. 2011a,b; Prantzos et al. 2018; Spitoni et al. 2018; Olive \& Vangioni 2019).
First, Timmes et al. (1995) showed that, in principle, the $\nu$-process could provide a possible explanation for the origin of fluorine, even if their yields of fluorine from core-collapse supernovae including the $\nu$-process were not enough to reproduce the observations.
Then, Meynet \& Arnould (2000) found that W-R stars can contribute significantly to the solar abundance of fluorine.
Moreover, Renda et al. (2004) took into account the contribution of both W-R stars and AGB stars to the evolution of fluorine, and concluded that W-R stars are fundamental to reproduce the fluorine abundance in the solar vicinity.
Spitoni et al. (2018) showed that the fluorine production is dominated by AGB stars, but the W-R stars are also required to reproduce the observations in the solar neighborhood.
Kobayashi et al. (2011a) found that, since the mass range of AGB stars that produces fluorine is 2-4 M$_{\odot}$, this contribution in Galactic chemical evolution models can be seen only at [Fe/H]$>$-1.5 dex, and it is not enough to reproduce the observations at [Fe/H]$\sim$0.
Furthermore, Kobayashi et al. (2011b) showed that both the $\nu$-process of core-collapse supernovae and the AGB stars can give a significant contribution to the production of fluorine: the main impact of the $\nu$-process in the [F/O] vs. [O/H] plot is represented by the presence of a plateau, followed by the rapid increase due to AGB stars.
In this context, Olive \& Vangioni (2019) showed that the $\nu$-process dominates at low metallicity, whereas the present-day fluorine abundance originates mainly from AGB stars. 
Moreover, Prantzos et al. (2018) took into account the yields from rotating massive stars by Limongi \& Chieffi (2018) in a Galactic chemical evolution model, and they showed that this process can dominate the fluorine production up to solar metallicities.
So far, also novae have been included in chemical evolution models following the evolution of fluorine; in particular, Spitoni et al. (2018) showed that novae can help to better reproduce the observed secondary behavior of fluorine in the [F/O] vs. [O/H] diagram.
\\Recently, several observational studies have appeared concerning fluorine in the Galaxy (e.g. Recio-Blanco et al. 2012; de Laverny \& Recio-Blanco 2013a,b; J{\"o}nsson et al. 2014, 2017; Pilachowski \& Pace 2015; Guer{\c{c}}o et al. 2019a,b; Ryde et al. 2020). In particular, Ryde et al. (2020) have provided stellar abundances of fluorine in a wide range of metallicity (-1.1$<$[Fe/H]$<$0.4) and for different stellar populations (both the thick and thin discs), and pointed out the need for several cosmic sources for fluorine. In fact, the observational data reflect various processes, that act on different timescales and thus in different ranges of metallicity; moreover, the observed trends show differences between the different stellar populations. In this context, the comparison with theoretical models is needed to further constrain the origin and evolution of fluorine.
\\The goal of this paper is thus to model the evolution of fluorine in the solar neighborhood, in the light of the recent data by Ryde et al. (2020). In particular, we adopt the reference models of Grisoni et al. (2017) for the Galactic thick and thin discs; these models have been constrained in order to reproduce the [$\alpha$/Fe] vs. [Fe/H] plots and the metallicity distribution functions (MDFs) (Grisoni et al. 2017) as well as the abundance patterns of different chemical elements such as lithium (Grisoni et al. 2019), carbon (Romano et al. 2020) and neutron capture elements (Grisoni et al. 2020), and now we apply them to study the evolution of fluorine.
\\This paper is organized in the following way. In Section 2, we describe the observational data considered in this work. In Section 3, we present the chemical evolution models adopted, with particular focus on the nucleosynthesis prescriptions for fluorine. In Section 4, we discuss our results based on the comparison between observational data and model predictions. Finally, in Section 5, we summarize our conclusions.

\section{Observational data}

The data discussed here is from Ryde et al. (2020). They determined the fluorine abundances in 61 giants from high-resolution and high signal-to-noise spectra of the HF molecular line at $2.3\,\mu$m.  These were observed with the Immersion GRating INfrared spectrograph (Yuk et al. 2010; Park et al. 2014) mounted on the 4.3-meter Lowell Discovery Telescope (LDT) and with the Phoenix  spectrograph (Hinkle et al. 2003) mounted on the 2.7 meter Harlan J. Smith Telescope at McDonald Observatory. The fluorine abundances were derived from fitting the molecular line with tailored synthetic spectra and stellar atmospheres models. The stellar parameters of the observed stars as well as the oxygen abundances presented in Ryde et al. (2020) were carefully and homogeneously determined from optical spectra minimising the systematic uncertainties in the derived  fluorine abundances inherent of the used HF line.  The uncertainties in the abundance ratios are of the order of 0.1 dex. However, for 10 of  the 20 stars observed with IGRINS, only upper limits were measured.

\subsection{Thick and thin discs}

The assigned stellar population of the stars, i.e. thin disc, thick disc, or halo is based on a hybrid approach using both abundances and kinematics on a much larger sample of stars with optical spectra (J\"onsson et al. in prep). 
For [Fe/H] between -0.9 to 0.15 the separation in thin and thick disc is solely based on the splitting in the [Mg/Fe] vs. [Fe/H] plane, and all stars with [Fe/H]$< -1.2$ are classified as halo stars (there are no such stars in the subsample of stars with determined fluorine abundances discussed in this paper).
For metallicities where it is hard to use the abundances for classification, kinematics are used instead. For these stars the space velocities (U, V, W and the total velocity V$_{tot}$) are calculated using radial velocities determined from the optical spectra, proper motions and positions from {\it Gaia} DR2 ({\it Gaia} collaboration, Prusti et al. 2016; Brown et al. 2018; Lindegren et al. 2018), and distances from McMillan (2018).
For stars with -1.2$<$[Fe/H]$<$-0.9, those with V$_{tot}>200$ km s$^{-1}$ are considered halo stars (there is one such star in the subsample used in this paper).
For stars with [Fe/H]$>$+0.15, those with V$_{tot}<70$ km s$^{-1}$ are considered thin disc stars.

\section{Chemical evolution models}

In this work, we use the parallel model, developed by Grisoni et al. (2017) (see also Grisoni et al. 2019, 2020). In this model, we assume that the thick and thin discs form by means of two separate infall episodes and they evolve at different rates. Thus, the evolution of the two components is disentangled and it is possible to follow separetely what happens in the thick and thin discs.
\\In this scenario, the gas infall rate laws for a certain element $i$ at the Galactocentric distance $r$ and time $t$ are given by:
\begin{equation} \label{eq_1IMthick}
(\dot G_i(r,t)_{inf})|_{thick}=A(r)(X_i)_{inf}e^{-\frac{t}{\tau_1}},
\end{equation}
and
\begin{equation} \label{eq_1IMthin}
(\dot G_i(r,t)_{inf})|_{thin}=B(r)(X_i)_{inf}e^{-\frac{t}{\tau_2}},
\end{equation}
for the Galactic thick disc and for the thin disc, respectively. The quantity $(X_i)_{inf}$ is the abundance by mass of the element $i$ in the infalling gas. The parameters $\tau_1$ and $\tau_2$ represent the timescales for mass accretion in the thick and thin disc components, respectively: they are free parameters of our model and they are constrained mainly by the comparison with the observed MDF of long-lived stars in the solar vicinity. In particular, $\tau_1$ is set equal to 0.5 Gyr, whereas $\tau_{2}(r)$ is 7 Gyr in the solar vicinity (Grisoni et al. 2019, 2020). The quantities $A(r)$ and $B(r)$ are two parameters fixed by reproducing the present time total surface mass density in the solar neighbourhood as given by Nesti \& Salucci (2013).
\\The star formation rate (SFR) is given by the Schmidt-Kennicutt law (Schmidt 1959; Kennicutt 1998a,b):
\begin{equation} \label{eq_03_02}
\psi(t) \propto \nu \sigma_{gas}^k,
\end{equation}
where $\sigma_{gas}$ is the surface gas density, k = 1.4 the law index, and $\nu$ the star formation efficiency ($\nu =$ 2 and 1 Gyr$^{-1}$ in the thick and thin disc components, respectively).
The initial mass function (IMF) is the Kroupa et al. (1993) one.
\\The prescriptions adopted here are the ones for the solar neighborhood as in Grisoni et al. (2017) (see also Grisoni et al. 2019, 2020), although the model has been extended to the other Galactocentric distances in Grisoni et al. (2018).

\subsection{Nucleosynthesis prescriptions}

The different nucleosynthesis prescriptions adopted in the models are summarized in Table 1, and in the following we describe them in details.

\subsubsection{Single stars}

Models labelled V300, V150 and V000 in Table 1 adopt the same nucleosynthesis prescriptions as models MWG-05, MWG-06, and MWG-07 of Romano et al. (2019), namely, the yields from Ventura et al. (2013, and private communication) for non-rotating low- and intermediate-mass stars (LIMS) as well as super-AGB stars, and the yields from Limongi \& Chieffi (2018, their recommended set R) for massive stars with initial rotational velocities of, respectively, 0, 150, and 300 km s$^{-1}$. Model V075 is added, in which we consider an intermediate value for the initial rotational velocity of massive stars of 75 km s$^{-1}$. We compute the corresponding yields by interpolating linearly in between the published grids for v$_{rot}$ = 0 and 150 km s$^{-1}$. However, we caution that this is a risky procedure, since published yields are not linear functions of v$_{rot}$. In other words, we by no mean intend to demonstrate that the average rotational velocity of massive stars in the early Galaxy must be 75 km s$^{-1}$. We only want to highlight that in the low-metallicity domain, say for [Fe/H]<-0.5, some intermediate value of the rotational velocity  should be adopted rather than the extreme ones considered by Limongi \& Chieffi (2018) in order to fit better the data.
Models Vvar and Kvar adopt the same nucleosynthesis prescriptions of Models MWG-11 and MWG-12 of Romano et al. (2019), respectively. Briefly, following the suggestions of Romano et al. (2019), we assume the yields for massive fast rotators (v$_{rot}$ = 300 km s$^{-1}$) of Limongi \& Chieffi (2018) for [Fe/H]<-1 and the yields for non-rotating massive stars by the same authors above such metallicity threshold. The yields for stars with masses in the range of 1-9 M$_{\odot}$ are either from Ventura et al. (2013, and private communication) -- model Vvar -- or Karakas (2010) and Doherty et al. (2014a,b) -- model Kvar. The reader is referred to Romano et al. (2019) for more details on the adopted stellar nucleosynthesis prescriptions as well as for chemical evolution model results regarding the CNO elements.
Model Vx05 is the same as model Vvar, but the yields of $^{19}$F from LIMS and super-AGB stars are multiplied by a factor of five. Finally, model Vnov is the same as model Vvar, but $^{19}$F production from novae is included (see next subsection), while it was neglected in all previous models.

\subsubsection{Binary systems}

\begin{table*}
\caption{Nucleosynthesis prescriptions for the models considered in this work. In column (1), there is the model name. In columns (2) and (3), the prescriptions for low- and intermediate-mass stars (LIMS) and the ones for super asymptotic giant branch stars (super-AGB). In column (4), the ones for massive stars with the corresponding rotational velocity reported in column (5). Finally, in column (6), we state if nova nucleosynthesis is, or is not, included in the model.}
\label{tab_01}
\begin{center}
\begin{tabular}{c|cccccccccc}
  \hline
\\
 Yield set & LIMS & Super-AGB & Massive stars & v$_{rot}$ (km s$^{-1}$) & Novae\\
 (1) & (2) & (3) & (4) & (5) & (6)\\
\\










 \hline

V300 & \multicolumn{2}{c}{Ventura et al. (2013) \& unpublished} & Limongi \& Chieffi (2018) & 300 & No \\

 \hline

V150 & \multicolumn{2}{c}{Ventura et al. (2013) \& unpublished} & Limongi \& Chieffi (2018) & 150 & No \\

 \hline

V075 & \multicolumn{2}{c}{Ventura et al. (2013) \& unpublished} & Limongi \& Chieffi (2018) & 75 & No \\

 \hline

V000 & \multicolumn{2}{c}{Ventura et al. (2013) \& unpublished} & Limongi \& Chieffi (2018) & 0 & No \\

 \hline

Vvar & \multicolumn{2}{c}{Ventura et al. (2013) \& unpublished} & Limongi \& Chieffi (2018) & Variable$^a$ & No \\

 \hline

Kvar & Karakas (2010) & Doherty et al. (2014a,b) & Limongi \& Chieffi (2018) & Variable$^a$ & No \\

 \hline

Vx05 & \multicolumn{2}{c}{Ventura et al. (2013) \& unpublished (x5)} & Limongi \& Chieffi (2018) & Variable$^a$ & No \\

 \hline

Vnov & \multicolumn{2}{c}{Ventura et al. (2013) \& unpublished} & Limongi \& Chieffi (2018) & Variable$^a$ & Yes \\

 \hline
\\
\end{tabular}

{\raggedright \textbf{Notes.} $^a$See Section 3.1.1, and Romano et al. (2019) for further details. \par}

\end{center}
\end{table*}

\begin{figure*}
\includegraphics[scale=0.53]{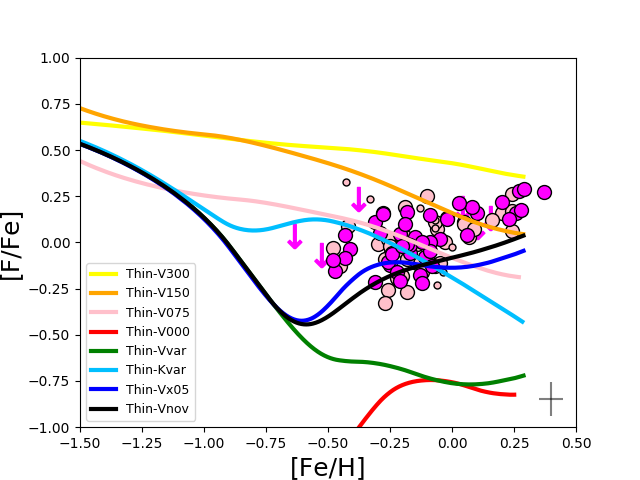}
\includegraphics[scale=0.53]{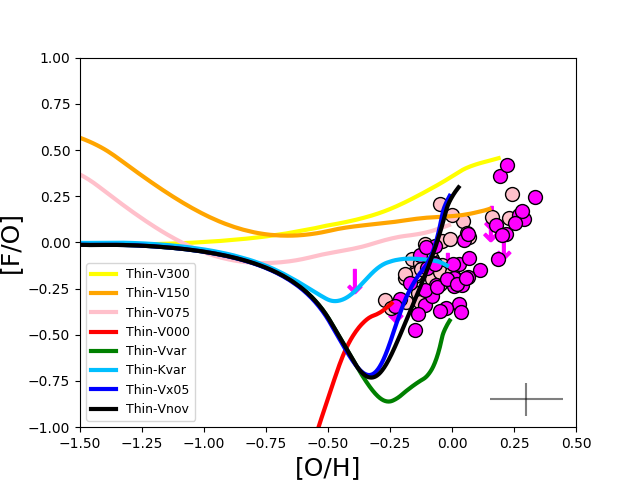}
 \caption{\textit{Left panel}: Observed and predicted [F/Fe] vs. [Fe/H] for the Galactic thin disc. The predictions are from the reference model for the Galactic thin disc, with the different nucleosynthesis prescriptions summarized in Table 1. The data for the Galactic thin disc are taken from Ryde et al. (2020) (magenta circles, the down arrows represent the corresponding upper limits), compared to determinations from literature (pink circles are from J{\"o}nsson et al. 2017, pink dots are from Guer{\c{c}}o et al. 2019b). \textit{Right panel}: Same as the left panel, but for [F/O] vs. [O/H]. Typical error bars from Ryde et al. (2020) are plotted in each panel.}
 \label{fig_01}
\end{figure*}

For binary systems giving rise to SNe Ia, we consider the single-degenerate scenario for their progenitors, i.e. a white dwarf (WD) plus a red giant companion (see Matteucci et al. 2009 and references therein). The adopted stellar yields for SNe Ia are those of Iwamoto et al. (1999).
\\As mentioned in the Introduction, also novae can in principle contribute to fluorine production (Jos{\'e} \& Hernanz 1998) and thus they are considered in this work. The progenitors of novae are binary systems of a WD and a low-mass main-sequence star (see Matteucci 2012 and references therein). Novae were first included in a detailed Galactic chemical evolution model by D'Antona \& Matteucci (1991) and their contribution to fluorine production was first studied in Spitoni et al. (2018). Here, for the nova nucleosynthesis, we adopt the same prescriptions as in  Spitoni et al. (2018):  yields by  Jos{\'e} \& Hernanz (1998) for nova outbursts in  CO and ONe WD with masses in the range between  0.8 M$_{\odot}$  and 1.35  M$_{\odot}$.
Spitoni et al. (2018) showed that the inclusion of those yields has a negligible effect on the chemical evolution of fluorine and hence, because of significant uncertainties in the fluorine yields, they consider the maximum yield by Jos{\'e} \& Hernanz (1998; model ONe7) related to ONe WD with masses of 1.35 M$_{\odot}$.
In Table 1, the model labeled Vnov adopts the nova contribution with the maximum yield for fluorine multiplied by a factor of 5 as in the best model of Spitoni et al. (2018,  their Figure 10).\\
\\We underline the fact that the nucleosynthesis prescriptions represent the largest uncertainity in Galactic chemical evolution models (C{\^o}t{\'e} et al. 2017). Thus, in the literature several Galactic chemical evolution studies have proposed corrections to explain the observational data (see for example Fran{\c{c}}ois et al. 2004, more recently Matteucci et al. 2020). In particular, concerning the corrections considered in this work for the stellar yields of fluorine, they have been already suggested by previous Galactic chemical evolution studies. For example, model Vnov includes the nova contribution with the maximum yield for fluorine by  Jos{\'e} \& Hernanz (1998) multiplied by a factor of 5, as in the best model by Spitoni et al. (2018).
Moreover, the need for an increase in the fluorine yields of AGB stars was also mentioned in Prantzos et al. (2018), where they suggested an increase by a factor of 2 to better explain the observations.
Here, we adopt the prescriptions by Ventura et al. (2013). These prescriptions for low-mass AGB stars could be underestimated since they do not include a s-process reaction network (P. Ventura, private communication). Thus, they might require a multiplying factor and, in particular, we consider a factor of 5 in our model Vx05.
Such corrections to the stellar yields then need to be confirmed by further theoretical stellar studies.
In this context, we remind that fluorine nucleosynthesis can be strongly affected by the uncertainities related to the nuclear cross sections (Lugaro et al. 2004; Cristallo et al. 2014).

\section{Results}

In this Section, we discuss our results, based on the comparison between model predictions and observational data. Firstly, we consider different sets of yields in the model of the thin disc, in order to establish the best nucleosynthesis prescriptions to explain the origin and evolution of fluorine. Then, we apply the reference models to study its evolution in both the Galactic thick and thin discs.

\subsection{Results for the thin disc}

\begin{figure*}
\includegraphics[scale=0.53]{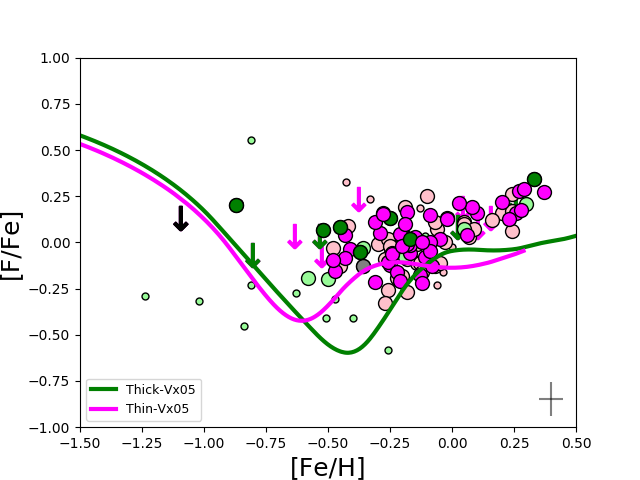}
\includegraphics[scale=0.53]{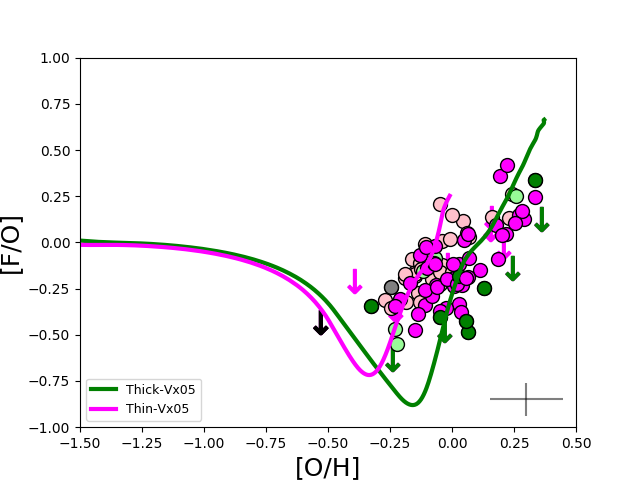}
\includegraphics[scale=0.53]{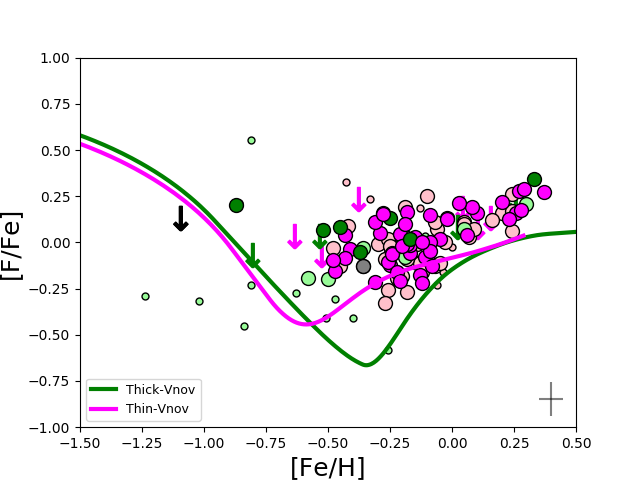}
\includegraphics[scale=0.53]{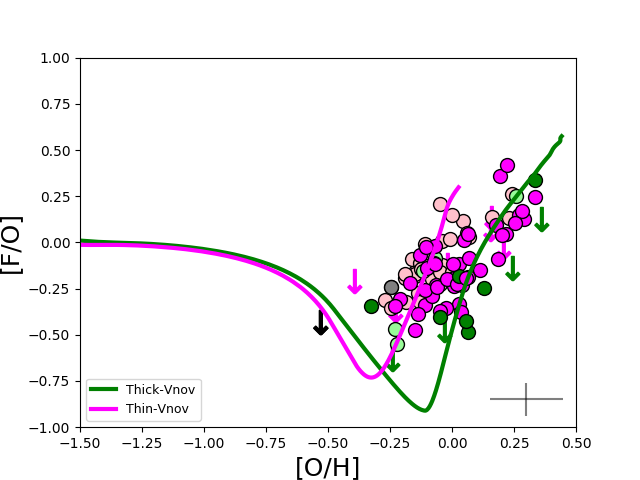}
 \caption{\textit{Upper left panel}: Observed and predicted [F/Fe] vs. [Fe/H] for the Galactic thick and thin discs. The predictions are from the reference models for the Galactic thick disc (green line) and thin disc (magenta line), in the case of the yield set Vx05. The data are taken from Ryde et al. (2020), and they are color-coded according to whether the stars belong to the thick disc (green circles) or thin disc (magenta circles) (the down arrows represent the corresponding upper limits -in black for the halo star- from Ryde et al. 2020), compared to determinations from literature (light-green and pink circles are for thick and thin disc stars -in gray a probable halo star- from J{\"o}nsson et al. 2017 ; light-green and pink dots are for thick and thin disc stars from Guer{\c{c}}o et al. 2019b). \textit{Upper right panel}: Same as the left panel, but for [F/O] vs. [O/H]. \textit{Lower panels}: same as the corresponding upper panels, but in the case of the yield set Vnov. Typical error bars from Ryde et al. (2020) are plotted in each panel.}
 \label{fig_02}
\end{figure*}

We start by testing different yield sets in the reference model of the Galactic thin disc; these yield sets are summarized in Table 1.
\\In the left panel of Fig. 1, we show the observed and predicted [F/Fe] vs. [Fe/H] for the Galactic thin disc. The predictions are from the reference model for the Galactic thin disc, compared to the recent data for thin disc stars by Ryde et al. (2020) (we also plot determinations for thin disc stars from literature by J{\"o}nsson et al. 2017 and Guer{\c{c}}o et al. 2019b).
The different predictions depend on the different sets of yields that have been implemented in the model (see Table 1). These sets of yields have already been included in Galactic chemical evolution models and tested to follow the evolution of CNO isotopes (see Romano et al. 2017, 2019, 2020).
As we can see, different sets of yields provide very different predictions for fluorine.
We note that the contribution from rapidly rotating massive stars can dominate the fluorine production up to solar metallicities.
Prantzos et al. (2018) first suggested the need for rotating massive stars at low metallicity. Moreover, they also proposed a distribution of rotational velocities in order to satisfy all the observational constraints and, in particular, to obtain the observed primary behaviour of nitrogen at low metallicities (requiring high rotational velocity) and to avoid overproduction of s-process elements at higher metallicities (requiring lower rotational velocity).
\\Here, we show that, in the case of rotational velocity 150 and 300 km s$^{-1}$, the fluorine production is dominated by the contribution of rotating massive stars, at variance with the non-rotating case. 
We also performed an intermediate test that we denote by 75 km s$^{-1}$, obtained by interpolation between the yield sets 0 and 150 km s$^{-1}$ (see Rizzuti et al. 2019). However, this is a risky procedure, since the published yields are not a linear function of the stellar initial rotational velocity.
A good assumption could be also a variable rotational velocity, with the most massive stars rotating fast during the earliest phases of Galactic evolution and much more slowly for [Fe/H]>-1, as suggested by Romano et al. (2019, 2020) on the basis of observations of CNO isotopes.
\\In general, the contribution of rotating massive stars can be important to explain the plateau at low metallicities, where different rotational velocities can set different values for the plateau, whereas the rise in the [F/Fe] vs. [Fe/H] diagram can be due to AGB stars which contribute at later times. Alternatively, also novae could help to explain the secondary behaviour, as suggested by Spitoni et al. (2018). Higher mass loss rates from metal-rich massive stars could also play a role.
\\In the right panel of Fig. 1, we use oxygen instead of iron as metallicity indicator, and we show the observed and predicted [F/O] vs. [O/H] diagram, widely used in literature to trace the evolution of fluorine.
Also in this case, different sets of yields can provide very different predictions for the thin disc and similar conclusions can be reached. In particular, the plateau in the [F/O] vs. [O/H] diagram at low [O/H] values can be obtained in the case with rotating massive stars at variance with the non-rotating case, and a good assumption to explain the overall behaviour could be the one with a variable rotational velocity, as previously mentioned. The rise in the [F/O] vs. [O/H] diagram at high [O/H] values would then be explained by the contribution of AGB stars or alternatively by the one of novae.

\subsection{Results for the thick disc}

Once the reference yield sets have been selected, we apply them to follow also the evolution of the Galactic thick disc.
\\The parallel model of Grisoni et al. (2017) allows us to follow separately the evolution of the thick and thin discs (see also Chiappini 2009). The chemical bimodality between the discs is evident in the [$\alpha$/Fe] vs. [Fe/H] plot (Grisoni et al. 2017, Spitoni et al. 2019), but also the abundance pattern of other chemical elements can be studied in this way and give important constraints on the different star formation histories of the two components, such as lithium (Grisoni et al. 2019), neutron-capture elements (Grisoni et al. 2020) and carbon (Romano et al. 2020). Here, we apply this approach to investigate fluorine.
\\In Fig. 2, we show the observed and predicted [F/Fe] vs. [Fe/H] and [F/O] vs. [O/H] for both the Galactic thick and thin disc. The predictions are from the parallel model of the thick and thin discs, compared to the recent data by Ryde et al. (2020) color-coded according to whether they belong to the thick or thin disc (we also plot determinations for thick and thin disc stars from literature by J{\"o}nsson et al. 2017 and Guer{\c{c}}o et al. 2019b).
Here, we show our predictions in the case of two different sets of yields; in particular, we include the contribution from rotating massive stars with variable rotational velocities, and consider the yield sets Vx05 and Vnov, as described in Table 1.
In both cases, we can see that the two sequences for the thin and thick discs seem to be explained, with the track of the thick disc shifted towards higher metallicities due to its faster evolution, in agreement with the so-called time-delay model (Tinsley 1980; Matteucci 2001, 2012; see also Kobayashi et al. 2011a concerning the time-delay effect in the abundance pattern of fluorine). In fact, the Galactic thick disc has a more intense star formation history than the thin disc. It is characterized by a much faster evolution, with a stronger star formation efficiency ($\nu$=2 Gyr$^{-1}$) and a shorter gas infall timescale ($\tau$=0.5 Gyr). Thus, it evolves more rapidly than the thin disc. The chemical evolution of the thick disc lasts for approximately 2 Gyr, and afterwards it shows negligible star formation. Hence, we predict a very low number of thick disc stars at higher metallicities, in agreement with the observed MDF of this component (Grisoni et al. 2017).
\\In particular, the dichotomy between the two discs is more evident in the [F/O] vs. [O/H] plot, where we clearly observe and predict the two sequences corresponding to the thick and thin discs. Similarly, in Romano et al. (2020) we found that the models fit better the [C/O] vs. [O/H] plane rather than the [C/Fe] vs. [Fe/H] and this might be due to the more uncertain yields of Fe.
\\In conclusion, we confirm that the thick disc has evolved much faster than the thin disc, in agreement with findings from other abundance patterns such as the $\alpha$-elements (Grisoni et al. 2017, Spitoni et al. 2019), lithium (Grisoni et al. 2019), neutron-capture elements (Grisoni et al. 2020) and carbon (Romano et al. 2020). 

\section{Conclusions}

In this paper, we have investigated the evolution of fluorine in the Milky Way thick and thin discs by means of detailed chemical evolution models and compared the model outputs with recent observational data. The main conclusions of our work can be summarized as follows.
\begin{itemize}
\item We investigate the contribution from rapidly rotating massive stars using the yields of Limongi \& Chieffi (2018) and we show that it can dominate the fluorine production up to solar metallicities, in agreement with Prantzos et al. (2018).
\item Different sets of yields with different rotational velocities can provide very different predictions in the [F/Fe] vs. [Fe/H] as well as [F/O] vs. [O/H] planes. In particular, the best agreement with the observations for fluorine is given by the assumption that most massive stars rotate fast during the earliest phases of Galactic evolution, while they rotate much more slowly or not at all at later times, as suggested by Romano et al. (2019, 2020) on the basis of data for CNO elements.
\item Other sites for the production of fluorine are required to explain the secondary behaviour at higher metallicities, such as AGB stars and/or novae.
\item Once the reference set of yields has been found, we apply it to study also the evolution of the Galactic thick disc. We confirm that this component has evolved much faster than the thin disc, in agreement with findings from other abundance patterns, such as the $\alpha$-elements (Grisoni et al. 2017, Spitoni et al. 2019), lithium (Grisoni et al. 2019), neutron-capture elements (Grisoni et al. 2020) and carbon (Romano et al. 2020).
\item In the case of fluorine, the dichotomy between the thick and thin discs seems to be more evident in the [F/O] vs. [O/H] plot, than in the [F/Fe] vs. [Fe/H] one. 
\end{itemize}
In conclusions, rotating massive stars can be important producers of fluorine. However, other producers are required to explain the secondary behaviour at higher metallicities, such as AGB stars and/or novae. Data for fluorine in other environments are needed to disentangle among the various hypotheses.

\section*{Acknowledgments}

V.G. acknowledges financial support at SISSA from the European Social Fund operational Programme 2014/2020 of the autonomous region Friuli Venezia Giulia.
E.S. acknowledges support from the Independent Research Fund Denmark (Research grant 7027-00096B). Funding for the Stellar Astrophysics Centre is provided by The Danish National Research Foundation (Grant agreement no.: DNRF106).
N.R. acknowledges support from the Royal Physiographic Society in Lund through  the Stiftelse Walter Gyllenbergs fond and M{\"a}rta och Erik Holmbergs donation and from the Crafoord Foundation.
H. J. acknowledges support from the Crafoord Foundation, Stiftelsen Olle Engkvist Byggm\"astare, and Ruth och Nils-Erik Stenb\"acks stiftelse.
We also thank Paolo Ventura for useful comments. Finally, we are grateful to an anonymous referee for the insightful comments and suggestions that improved this work.

\section*{Data availability}

The derived data generated in this research will be shared on reasonable request to the corresponding author.

\end{document}